\documentclass[a4paper,fleqn]{cas-sc}

\usepackage[numbers]{natbib}
\usepackage{setspace}
\usepackage{lineno}
\usepackage{comment}

\def\tsc#1{\csdef{#1}{\textsc{\lowercase{#1}}\xspace}}
\tsc{WGM}
\tsc{QE}
\tsc{EP}
\tsc{PMS}
\tsc{BEC}
\tsc{DE}

\begin{document}
\let\WriteBookmarks\relax
\def\floatpagepagefraction{1}
\def\textpagefraction{.001}
\shorttitle{Deterministic coherence and anti-coherence resonances}
\shortauthors{P.S.~Komkov et al.}

\title [mode = title]{Deterministic coherence and anti-coherence resonances in two coupled Lorenz oscillators: numerical study versus experiment}                      

\author[1]{Pavel S. Komkov}[orcid=0009-0002-5988-5352]
\ead[url]{k-pavlik-k@mail.ru}
\address[1]{Institute of Physics, Saratov State University, 83 Astrakhanskaya str., 410012 Saratov, Russia}

\author[1]{Ol'ga I. Moskalenko}[orcid=0000-0001-5727-5169]
\ead[url]{o.i.moskalenko@gmail.com}

\author[1]{Vladimir V. Semenov}[orcid=0000-0002-4534-8065]
\ead[url]{semenov.v.v.ssu@gmail.com}
\corref{cor1}
\cortext[cor1]{Corresponding author}

\author[1]{Sergei V. Grishin}[orcid=0000-0002-3654-3299]
\ead[url]{sergrsh@yandex.ru}

\begin{abstract}
We demonstrate the deterministic coherence and anti-coherence resonance phenomena in two coupled identical chaotic Lorenz oscillators. Both effects are found to occur simultaneously when varying the coupling strength. In particular, the occurrence of deterministic coherence resonance is revealed by analysing time realizations $x(t)$ and $y(t)$ of both oscillators, whereas the anti-coherence resonance is identified when considering oscillations $z(t)$ at the same parameter values. Both resonances are observed when the coupling strength does not exceed a threshold value corresponding to complete synchronization of the interacting chaotic oscillators. In such a case, the coupled oscillators exhibit the hyperchaotic dynamics associated with the on-off intermittency. The highlighted effects are studied in numerical simulations and confirmed in physical experiments, showing an excellent correspondence and disclosing thereby the robustness of the observed phenomena.
\end{abstract}



\maketitle

\doublespacing

\section{\label{sec:sec1}Introduction}
The essence of coherence resonance (CR) consists in growth of the noise-induced oscillation regularity when increasing the noise intensity in some range such that there exists an optimum value of the noise level corresponding to the most coherent oscillations. Such effects are observed in a broad variety of dynamical systems including excitable oscillators subject to stochastic forcing \cite{pikovsky,lindner,lee1998,lindner1999,semenov2017,bogatenko2018,korneev2024} and the non-excitable ones with a subcritical Hopf bifurcation \cite{ushakov2005,zakharova2010,semenov2015}. CR has also been found for single noise-driven chaotic oscillators where its mechanism is explained by switching between attractors \cite{palenzuela,calvo1,calvo2,komkov2025}, as well as in coupled chaotic systems \cite{kiss2003,zhan2002,zhou2002,liu}. One of the simplest model of coupled chaotic oscillators realizing CR represents two bidirectionally coupled identical or slightly non-identical Lorenz oscillators operating either in the regime of complete chaotic synchronization or near its threshold~\cite{liu}. In the presence of noise, the coupled chaotic systems exhibit the noise-induced on-off intermittency such that random switching between two distinct states (synchronized and asynchronized) occur. Transitions between on and off states here are interpreted as the motion near the fixed point and the excursion away from it, respectively, in excitable systems. Thus, qualitatively, CR in coupled chaotic systems and single excitable oscillators can be considered as related phenomena. 

A variety of dynamical systems exhibiting CR is not restricted to stochastic systems. A similar resonant effects can be realized in deterministic chaotic oscillators, where chaotic oscillations are considered as noise in stochastic dynamical systems. This phenomenon called deterministic coherence resonance (DCR) is found in single ~\cite{avila,bosco,grishin} and coupled ~\cite{pisarchik1,garcia,pisarchik2} chaotic oscillators. In such a case, the chaotic oscillations become more or less regular when varying the oscillators' parameters or the coupling strength similarly to the noise intensity in classical CR. When the DCR occurs, the oscillation regularity first increases and then decreases. Resultantly, one can distinguish the most regular oscillations for appropriate parameter values. 

The opposite process characterises the phenomenon of deterministic anti-coherence resonance (DACR) found to be exhibited by a network of coupled R{\"o}ssler oscillators \cite{jaimes}. In this scenario, decreasing the oscillation regularity is observed earlier than growth when changing the coupling strength. Consequently, there is a certain range of the coupling strength corresponding to the least coherent oscillations. As reported in Ref.~\cite{jaimes}, the reason for emergence of the DACR is a small mismatch between the natural frequencies of the R{\"o}ssler oscillators networked unidirectionally in a star-ring configuration. 

In the current paper, we extend a manifold of effects related to DCR and DACR by considering one more system, two bidirectionally coupled identical Lorenz oscillators. We combine methods of numerical simulation and experimental research by using an electronic model of the coupled Lorenz oscillators.
The system under study is assumed to be one of the simplest model for implementing DCR and DACR. Indeed, the considered model does not involve the frequency mismatch and exclude from the consideration the impact of the coupling topology. In addition, we demonstrate that DCR and DACR can be simultaneously exhibited by the same oscillators when the coupling strength growths. In particular, both effects are manifested when analyzing time realizations of different dynamical variables at the same parameter values.

\section{\label{sec:2} Model and methods}
Both DCR and DACR are explored in the present research on an example of two bidirectionally coupled identical Lorenz oscillators: 
\begin{equation}
\label{eq:1} 
\begin{array}{l} 
\dfrac{dx_{1,2}}{dt} = \sigma(y_{1,2}-x_{1,2})+K(x_{2,1}-x_{1,2}),\\
\dfrac{dy_{1,2}}{dt} = x_{1,2}(\rho-z_{1,2})-y_{1,2},\\
\dfrac{dz_{1,2}}{dt} = -\beta z_{1,2}+x_{1,2}y_{1,2},
\end{array}
\end{equation}
where $\sigma=10$, $\rho=28$ and $\beta=8/3$ are the oscillators' parameters assumed to be fixed. For chosen set of the parameter values, the coupling-free oscillators (see Eqs. (\ref{eq:1}) at $K \equiv 0$) exhibit the chaotic dynamics. In contrast to $\sigma$, $\rho$ and $\beta$, the coupling strength $K$ plays a role of a control parameter and varies from $0$ to $10$. Our investigations are performed by means of numerical simulations and electronic experiments. In more detail, we integrate Eqs. (\ref{eq:1}) numerically using the fourth-order Runge-Kutta method with time step $\Delta t=0.005$. The used initial conditions are chosen to be random and uniformly distributed in the ranges $x_{1,2}(t=0)\in [-1,1]$, $y_{1,2}(t=0)\in [-1,1]$, $z_{1,2}(t=0)\in [-1,1]$. The total integration time is $t_{\text{total}}=10^5$. Thus, numerically obtained realizations $x_{1,2}(t)$, $y_{1,2}(t)$ and $z_{1,2}(t)$ of length $t_{\text{total}}/\Delta t$ are used for further time series analysis.

\begin{figure}[t]
\centering
\includegraphics[width=0.85\textwidth]{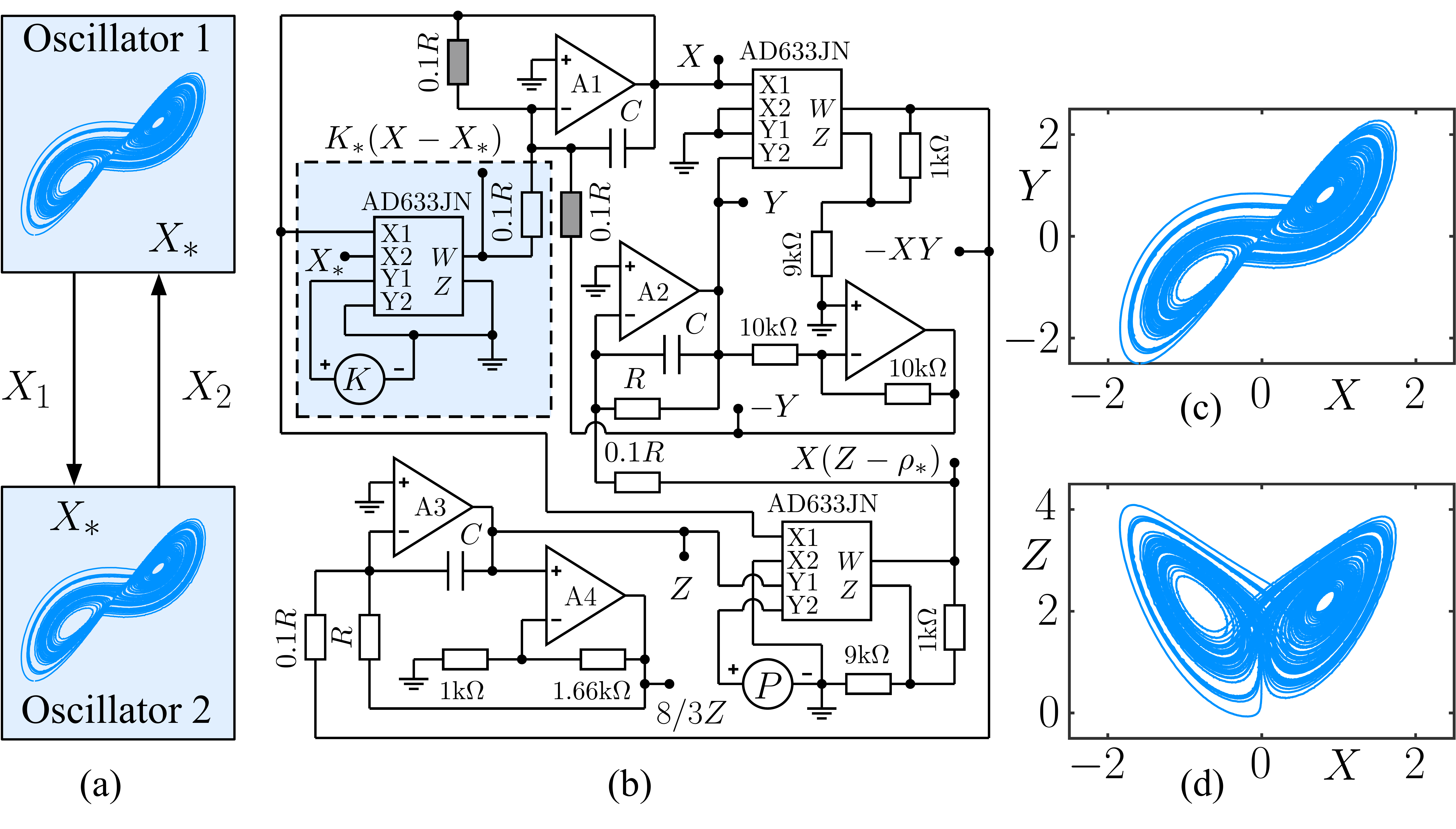}
\caption{Electronic model of two bidirectionally coupled chaotic Lorenz oscillators (see Eqs.(\ref{eq:2})): (a) Schematic illustration of the experimental setup where the oscillators' attractors are illustrated by projections in phase plane ($X$,$Y$); (b) Electronic circuit of each oscillator (both oscillators are assumed to be identical).  Operational amplifiers are TL072CP. Analog integrator elements are $C=100$ nF and $R=10$ k$\Omega$; (c)-(d) Projections of the experimentally obtained single oscillator phase portrait (see Eqs. (\ref{eq:2}) at $K\equiv0$, $\sigma=10$, $\beta=8/3$ and $P=2.3$) in planes ($X$,$Y$) (panel (c)) and ($X$,$Z$) (panel (d)). Since oscillators 1 and 2 are assumed to be identical, indexes 1 and 2 in panels (c) and (d) are not specified.}
\label{fig1}
\end{figure}

For physical experiments, we have developed an experimental prototype (schematically illustrated in Fig.~\ref{fig1}~(a)) being an electronic model of system (\ref{eq:1}) implemented by principles of analog modelling \cite{luchinsky1998,semenov2024_book}. Figure~\ref{fig1}~(b) describes the circuit diagram of each oscillator, which contains three integrators, A1, A2 and A3, whose output voltages are taken as the dynamical variables, $X$, $Y$ and $Z$, respectively. All the input signals of the integrators are voltages designated on the circuit in Fig.~\ref{fig1}~(b) for the reader's convenience. One of them is the signal $K(X-X_{*})$ produced by the circuit block (coloured in light blue in Fig.~\ref{fig1}~(b)) being responsible for the action of external force $X_*$ on the oscillators.  Coupling is organized such that $X_*$ are signals $X_1(t)$ and $X_2(t)$ acting on oscillators 2 and 1 (lower and upper oscillators in Fig.~\ref{fig1}~(a)), respectively. Operation of the experimental setup is described by the following equations:
\begin{equation}
\label{eq:2} 
\begin{array}{l} 
RC\dfrac{dX_{1,2}}{dt} = \sigma(Y_{1,2}-X_{1,2})+K(X_{2,1}-X_{1,2}),\\
RC\dfrac{dY_{1,2}}{dt} = 10X_{1,2}(P-Z_{1,2})-Y_{1,2},\\
RC\dfrac{dZ_{1,2}}{dt} = -\beta Z_{1,2}+10X_{1,2}Y_{1,2},
\end{array}
\end{equation}
where $C=100$ nF, $R=10$ k$\Omega$ are the capacitances and resistances at the integrators A1, A2 and A3. Parameters $\sigma=10$ and $\beta=8/3$ are fixed, since their values are determined by the corresponding changeless resistances. In contrast, coefficients $K$ and $P$ are values of DC voltages applied as an input signal of analog multipliers AD633JN. This approach for specifying the coupling strength and the main parameter allows to instantaneously vary $K$ and $P$ in both oscillators and to guarantee obeying the equalities $K_1=K_2=K$ and $P_1=P_2=P$ (the same DC voltage sources are used for tuning $K_{1,2}$ and $P_{1,2}$). In the following, parameter $P$ is assumed to be fixed, $P=2.3$ (corresponds to the chaotic dynamics of interacting oscillators, see Fig. \ref{fig1}~(c),(d)), whereas the coupling strength is varied in range [$0:10$ V]. For the reader's convenience, all the parameters and dynamical variables of Eqs. (\ref{eq:2}), their units and brief descriptions are summarized in table (\ref{table1}). 
\begin{table}[h!]
\begin{tabular}{| p{0.02\textwidth} | p{0.11\textwidth} | p{0.11\textwidth} | p{0.25\textwidth} |p{0.28\textwidth} |}
\hline
 & Fixed value or allowable range of values & Units & Description & Transformation into dimensionless model (\ref{eq:1}) \\
\hline
$R$ & 10 & k$\Omega$ & Reference resistance value of integrators A1-A3, determines the circuit's time constant & Introduction of dimensionless time as  $t=t/RC$\\ 
\hline
$C$ & 100 & nF & Capacitance of the capacitors at integrators A1-A3, determines the circuit's time constant & Introduction of dimensionless time as  $t=t/RC$\\ 
\hline
$X$ $Y$ $Z$ & [-10:10] & V & Output signals (voltages) of integrators A1-A3, describe dynamical variable oscillations & $x_{1,2}=10X_{1,2}/V_0$, $y_{1,2}=10Y_{1,2}/V_0$, $z_{1,2}=10Z_{1,2}/V_0$, where $V_0=1$ V \\ 
\hline
$K$ $P$ & [-10:10] & V & DC voltages which determine values of the corresponding parameters, inputs of analog multipliers & $K=K/ V_0$, $\rho=10 P/V_0$, where $V_0=1$ V \\ 
\hline
$\beta$ & 8/3 & dimensionless & Closed-loop gain of non-inverting amplifier A4 & No rescalling procedures are applied\\ 
\hline
$\sigma$ & 10 & dimensionless & Ratio of integrator's A1 input resistances (coloured in grey) to the reference resistance value $R$, equals to $R/0.1R$. & No rescalling procedures are applied \\ 
\hline
\end{tabular}
\caption{Brief overview on parameters and dynamical variables of Eqs. (\ref{eq:2}) referring to Fig. \ref{fig1}.}
\label{table1}
\end{table}

Model of two oscillators governed by Eqs. (\ref{eq:2}) can be transformed into dimensionless model (\ref{eq:1}) by using substitution $t=t/\tau_0$ ($\tau_0=RC=1$ ms is the circuit's time constant) with $\sigma=10$, $\rho=10 P$ and dynamical variables $x_{1,2}=10X_{1,2}/V_0$, $y_{1,2}=10Y_{1,2}/V_0$ and $z_{1,2}=10Z_{1,2}/V_0$, where $V_0$ is the unity voltage, $V_0=1$ V. Experimentally obtained time series $X_{1,2}$, $Y_{1,2}$, $Z_{1,2}$ were recorded from the corresponding outputs (marked in Fig.~\ref{fig1}~(b) as $X$, $Y$, $Z$) using an acquisition board (National Instruments NI-PCI 6133). All the experimental signals were digitized at the sampling frequency of 400 kHz (Fig.~\ref{fig1}~(c),(d) and Fig.~\ref{fig4}) and 50kHz (Fig.~\ref{fig5}~(c),(d) and Fig.~\ref{fig6}~(b)). 60 s long realizations were used for further offline processing whose results are depicted in Fig.~\ref{fig5}~(c),(d) and Fig.~\ref{fig6}~(b).

Similarly to mathematical model (\ref{eq:1}), experimental oscillators (\ref{eq:2}) are assumed to be identical, but in fact a parameter mismatch inevitably takes place due to deviations of real capacitances and resistances from their nominal values. Following this feature, all the resistors and capacitors were chosen such that the deviation from the nominal values does not exceed 2 percent. Secondly, the used operational amplifiers and analog multipliers are characterised by small, but nonzero output voltage offset, which is an unique characteristic for each integrated circuit. There are additional factors such as measurement inaccuracies, slight internal fluctuations, electronic component imperfectness (for instance, capacitors’ leakage currents), finite bandwidth of operation amplifiers and analog multipliers, etc. Individually, all the factors and effects mentioned above very slightly impact the oscillatory dynamics. However, the total effect of all the inaccuracies and imperfectnesses results in the inability to achieve complete synchronization within the available coupling strength range (see Sec. \ref{sec:3}). Nevertheless, very similar statistical properties of oscillations in the on-off intermittency regime were established in numerical and physical experiments. Moreover, the DCR and DACR occur in a similar manner when increasing the coupling strength in mathematical model (\ref{eq:1}) and experimental setup (\ref{eq:2}).

To reveal the intrinsic properties of the coupled oscillator dynamics, we explore the evolution of the time realizations, phase portraits and Lyapunov exponent spectrum caused by the coupling strength growth. To characterise the on-off intermittency, two statistical characteristics are analyzed: the distribution of laminar phase lengths $N(\tau)$ and the mean laminar phase length $\langle \tau \rangle$.

In addition, we consider the correlation time, $t_{\text{cor}}$, to describe the DCR and DACR similarly to classical coherence resonance. The correlation time is introduced in the following form:
\begin{equation}
\label{eq:t_cor} 
t_{\text{cor}}=\dfrac{1}{\Psi(0)}\int\limits_{0}^{\infty} \left| \Psi(s) \right|ds,
\end{equation} 
where $\Psi(s)$ and $\Psi(0)$ are the autocorrelation function and the variance of the time realizations $x_{1,2}(t)$, $y_{1,2}(t)$ and $z_{1,2}(t)$. In the following, the evolution of the dynamics caused by increasing the coupling strength $K$ is described by using dependencies of $t_{\text{cor}}(K)$. Mathematical model (\ref{eq:1}) and experimental setup equations (\ref{eq:2}) have different time scales. For this reason, the correlation times registered in numerical and physical experiments differ by $RC$ times. To present the correlation time change in the same scale, the dependence of a normalized correlation time $\tilde{t}_{\text{cor}}$ on the coupling strength $K$ is taken into consideration, $\tilde{t}_{\text{cor}}(K)=t_{\text{cor}}(K)/t_{\text{cor}}(K=0)$, where $t_{\text{cor}}(K=0)$ characterises the coupling-free dynamics.

Finally, the transformation of the normalized power spectral density (NPSD) when varying the coupling strength is included into consideration. First, the power spectral density, PSD($f$), of oscillations $x_1(t)$, $y_1(t)$ and $z_1(t)$ is calculated. After that, the NPSD is extracted as NPSD(f)=PSD($f$)/max(PSD($f$)) and expressed in decibels. The same procedure for the NPSD calculation is used in numerical simulations and physical experiments. Since the power spectra of oscillations $x_1(t)$ and $x_2(t)$, $y_1(t)$ and $y_2(t)$, $z_1(t)$ and $z_2(t)$ are identical, only NPSD of processes $x_1(t)$, $y_1(t)$ and $z_1(t)$ are presented in the paper.

\section{\label{sec:3} Results}
\subsection{\label{sec:3.1} On-off intermittency}
\begin{figure}[t]
\centering
\includegraphics[width=0.85\textwidth]{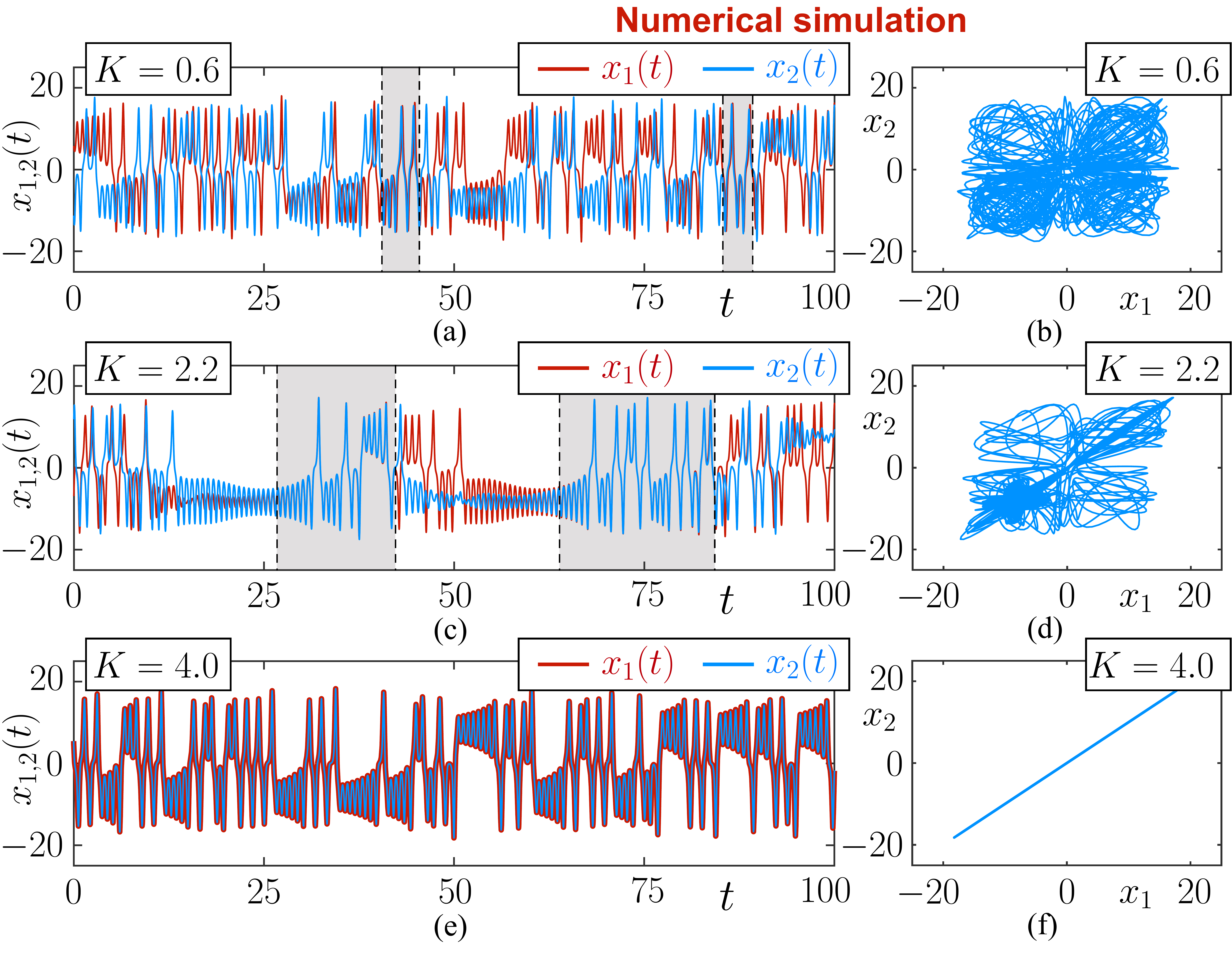}
\caption{Evolution of the oscillatory dynamics in numerical model (\ref{eq:1}) when increasing the coupling strength: $K=0.6$ (panels (a) and (b)), $K=2.2$ (panels (c) and (d)) and $K=4.0$ (panels (e) and (f)).  The dynamics evolution is illustrated by using time realizations $x_{1,2}(t)$ (left panels) and trajectories in phase plane ($x_{1}$, $x_2$) (right panels).  The time periods corresponding to the laminar phase in panels (a) and (c) are coloured in grey. The oscillators' parameters are $\sigma=10$, $\rho=28$, $\beta=8/3$. Random initial conditions within the range [-1:1] are used.}
\label{fig2}
\end{figure}  
\begin{figure}[t]
\centering
\includegraphics[width=0.85\textwidth]{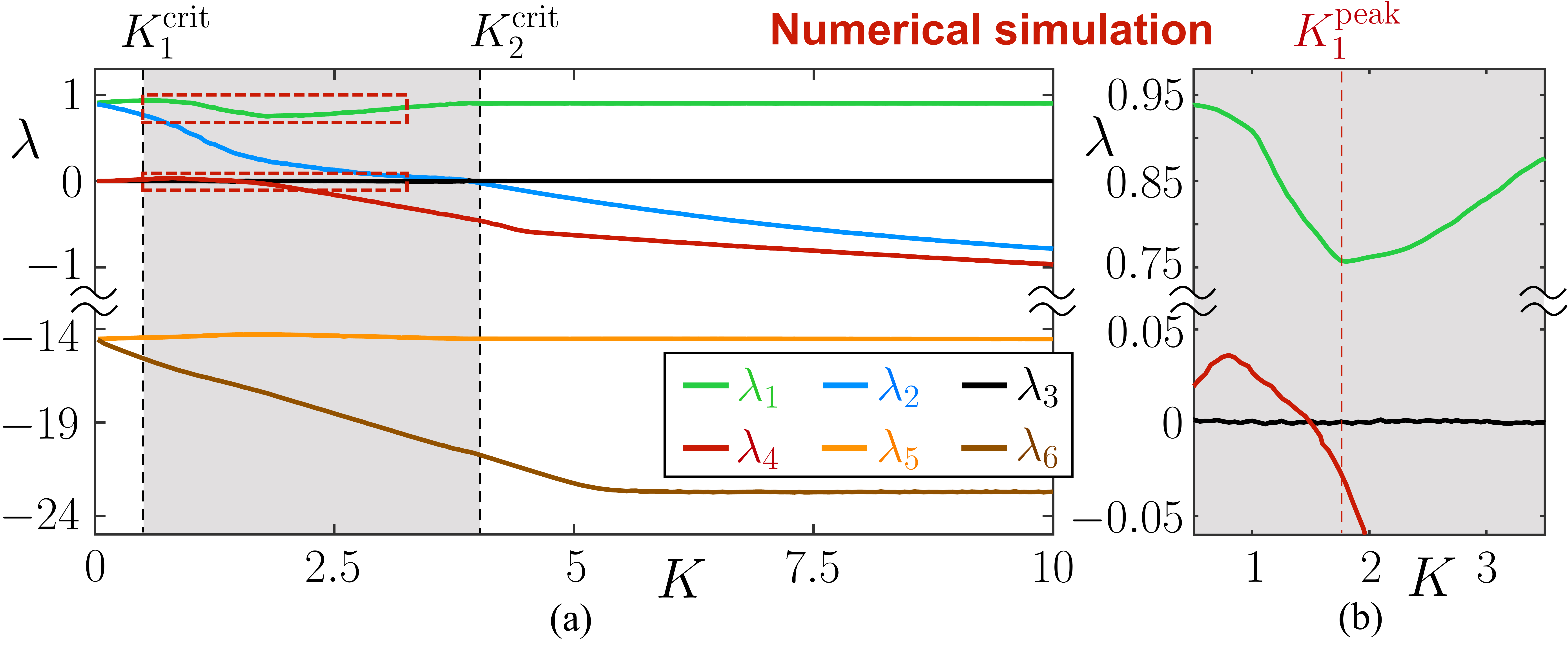}
\caption{Evolution of the Lyapunov exponent spectrum in numerical model (\ref{eq:1}): panel (a) illustrates changes caused by increasing the coupling strength in range $K\in [0:10]$, panel (b) describes the dependencies $\lambda_{1,3,4}(K)$ in the areas delineated by the red dashed rectangles in panel (a). The range $K\in(K^{\text{crit}}_{1}:K^{\text{crit}}_{2})$ corresponds to the occurrence of the on-off intermittency and is coloured in grey. The oscillators' parameters are $\sigma=10$, $\rho=28$, $\beta=8/3$. Random initial conditions within the range [-1:1] are used.}
\label{fig3}
\end{figure}  
Increasing the coupling strength in system (\ref{eq:1}) expectedly induces transition to the synchronous dynamics. However, the transition to synchronization is characterized by intrinsic peculiarities. In the presence of weak coupling, the behaviour of coupled oscillators does not principally differ from the single oscillator dynamics and the effect of synchronization is not observed. After the coupling strength passes through the critical value $K^{\text{crit}}_1\approx 0.5$, the asynchronous behaviour occasionally transforms into the synchronous dynamics (marked by the grey areas in Fig. \ref{fig2}~(a),(c)) and back to the asynchronous one. These spontaneous transitions are qualitatively equivalent to switches between laminar (synchronous behaviour) and turbulent (asynchronous dynamics) phases characterizing the on-off intermittency in chaotic systems. 
Further growth of the coupling strength leads to an increase in the duration of the synchronous states and in the frequency of their occurrence (compare Fig. \ref{fig2}~(a) and Fig. \ref{fig2}~(c)). The observed transformation culminates in arising complete synchronization of chaotic oscillations when passing through the second threshold value $K=K^{\text{crit}}_2\approx3.92$. After one achieves complete synchronization, interacting oscillators exhibit identical oscillations as demonstrated in Fig. \ref{fig2}~(e). Thus, one can characterise the area of the on-off intermittency occurrence as $K\in(K^{\text{crit}}_1:K^{\text{crit}}_2)$ (the grey area in Fig.\ref{fig3} and Fig.\ref{fig6}). Continuous character of the dynamics evolution  caused by coupling strength growth is also reflected in the phase portraits in plane ($x_1:x_2$) which gradually transform into the line $x_1=x_2$ [Fig. \ref{fig2}~(b),(d),(f)]. 

The oscillatory dynamics transformation in numerical model (\ref{eq:1}) described above is partially reflected in the evolution of the Lyapunov exponent spectrum when increasing the coupling strength [Fig. \ref{fig3}]. In particular, when the transition to complete synchronization at $K=K_{2}^{\text{crit}}$ occurs, Lyapunov exponent $\lambda_2$ becomes negative.  After passing through the threshold value $K=K^{\text{crit}}_2$, increasing the coupling strength does not induce qualitative changes in the dynamics. In particular, the oscillatory dynamics of model (\ref{eq:1}) at $K>K^{\text{crit}}_2$ is characterised by the only positive Lyapunov exponent $\lambda_1$ possessing a constant value. In contrast, two distinguishable effects occur at lower values of $K$. Namely, one of the Lyapunov exponents (see $\lambda_4$ in Fig. \ref{fig3}) becomes negative at $K\approx 1.49$. Secondly, Lyapunov exponent $\lambda_1$ is found to be non-monotonic such that there exist a local minimum at $K\approx 1.8$ marked as $K_1^{\text{peak}}$ in Fig.~\ref{fig3}~(b). Two revealed effects have no visible impact on the system dynamics when considering time realizations and phase portraits. However, as will be shown below, the non-monotonic behaviour of $\lambda_1$ correlates with DCR (see the next section). 

\begin{figure}[t]
\centering
\includegraphics[width=0.85\textwidth]{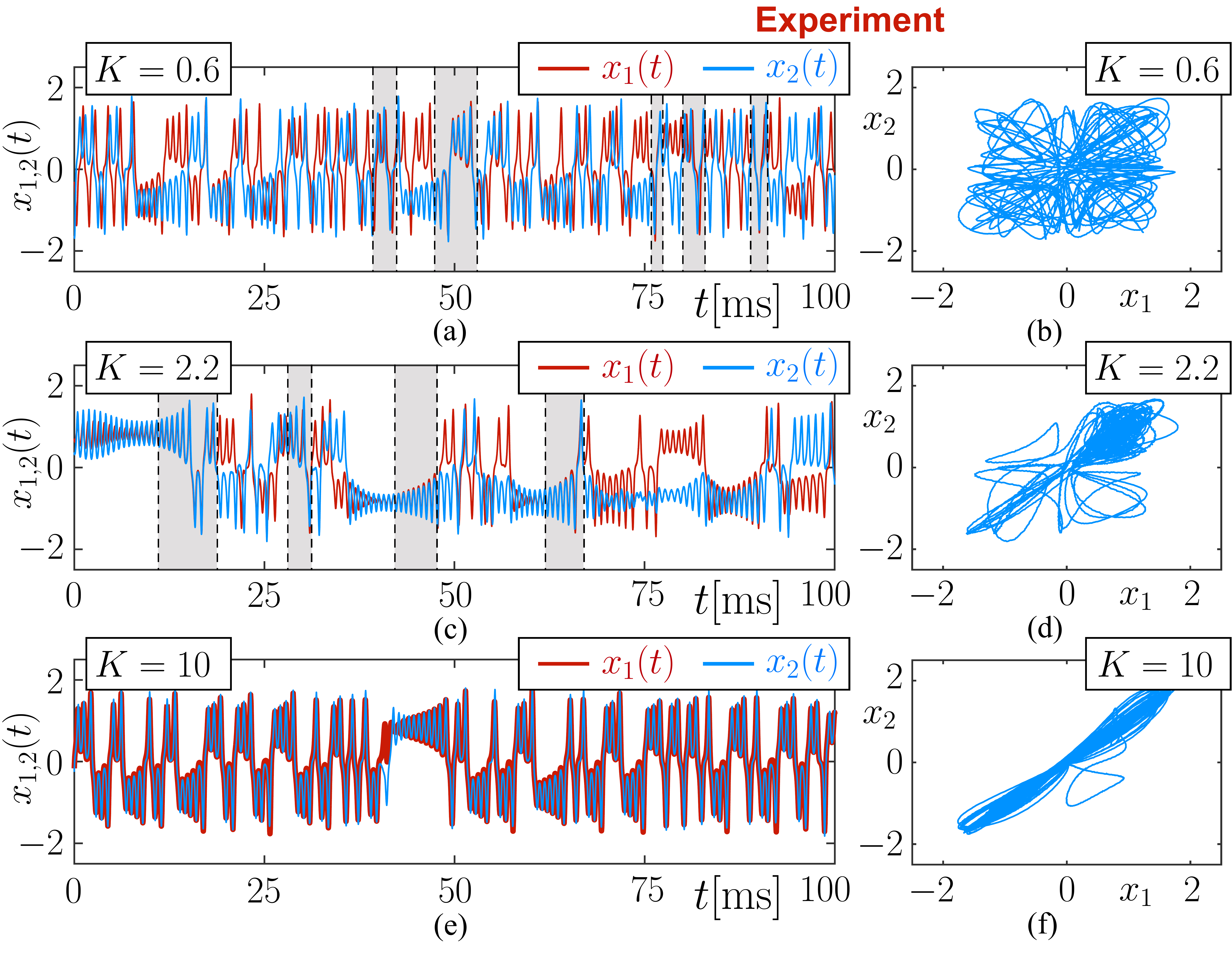}
\caption{Evolution of the oscillatory regimes in electronic setup (\ref{eq:2}) when increasing the coupling strength: $K=0.6$ (panels (a) and (b)), $K=2.2$ (panels (c) and (d)) and $K=10.0$ (panels (e) and (f)). The dynamics evolution is illustrated by using time realizations $x_{1,2}(t)$ and trajectories in phase plane ($x_{1}$, $x_2$) similarly to Fig.~\ref{fig2}. The time periods corresponding to the laminar phase in panels (a) and (c) are coloured in grey. The oscillators' parameters are $\sigma=10$, $P=2.3$ and $\beta=8/3$.}
\label{fig4}
\end{figure}

Figures \ref{fig2} and \ref{fig3} illustrate the effects observed in numerical model (\ref{eq:1}). Transitions 'asynchronous dynamics -- on-off intermittency' occur in experimental setup (\ref{eq:2}) in a similar way such that the temporary synchronized states begin to appear when the coupling strength becomes larger than the critical value $K_1^{\text{crit}}$. Further growth of the coupling strength makes such states more frequent and more longer (compare Fig.\ref{fig2}~(a)-(d) and Fig.\ref{fig4}~(a)-(d)). However, the transition to complete synchronization is not realized in electronic circuit (\ref{eq:2}) in the available range of the coupling strength, $K\in [0:10]$ (see Fig. \ref{fig4}~(e)-(f)), and is expected to potentially occur at higher coupling strengths which exceed the upper limit of 10. This is due to several factors. In particular, two chaotic oscillators (\ref{eq:2}) are in fact non-identical, their operation is characterised by the presence of inaccuracies, internal fluctuations and other factors inevitably presenting in real physical systems. As a result, complete synchronization of coupled chaotic oscillators is found to require higher values of the coupling strength as compared to numerical model (\ref{eq:1}).

\begin{figure}[t]
\centering
\includegraphics[width=0.85\textwidth]{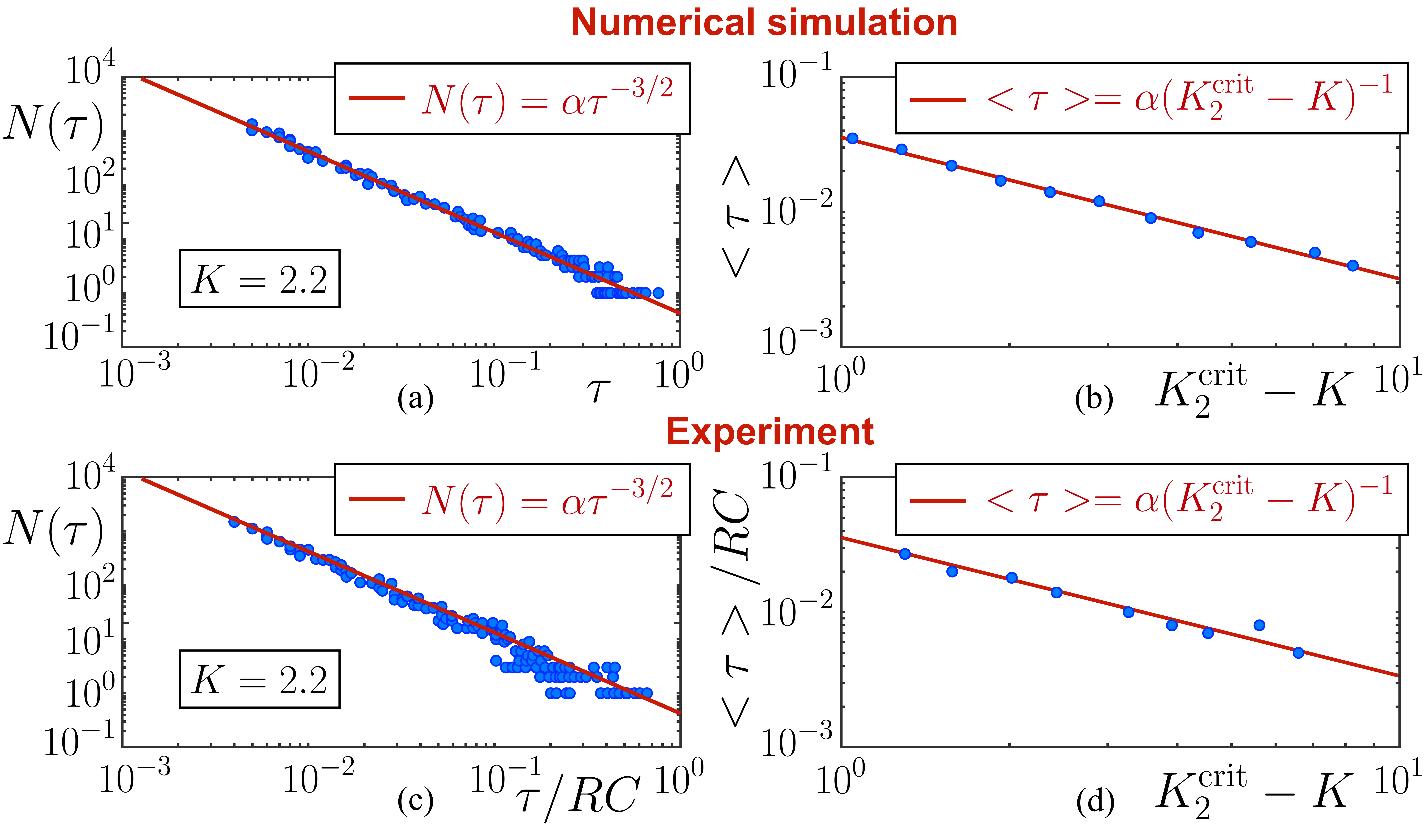}
\caption{Statistical characteristics of the on-off intermittency exhibited by numerical model (\ref{eq:1}) (panels (a) and (b)) and electronic setup (\ref{eq:2}) (panels (c) and (d)). Panels (a) and (c) illustrate the distribution of laminar phase lengths $N(\tau)$ at fixed coupling strength $K=2.2$, whereas panels (b) and (d) depict the dependencies of the mean laminar phase duration $<\tau>$ on the critical onset parameter $K_2^{\text{crit}}-K$. Since the mathematical model and experimental setup have different timescales, $\tau$ and $<\tau>$ in panels (c) and (d) are rescalled by $(RC)^{-1}$. Panels (a)-(d) contain red solid lines which represent the results of curve-fitting using the functions noted in the legends. The parameters estimated by means of curve-fitting are: $\alpha=0.4216$ (panel (a)), $\alpha=0.03663$ (panel (b)), $\alpha=0.2902$ (panel (c)) and $\alpha=0.03566$ (panel (d)). The parameter values of Eqs. (\ref{eq:1}) and (\ref{eq:2}) are $\sigma=10$, $\beta=8/3$, $\rho=28$ (numerical simulation), $P=2.3$ (physical experiment). Random initial conditions within the range [-1:1] are used for numerical simulations.}
\label{fig5}
\end{figure}

Despite the revealed difference between numerical model (\ref{eq:1}) and electronic setup (\ref{eq:2}), the on-off intermittency explored by means of numerical simulation and physical experiments occurs in the same way and is characterised by similar statistics as compared to the classical on-off intermittency. To confirm this fact, let us consider the distribution of the laminar phase lengths, $N(\tau)$, obtained in numerical and physical experiments at fixed coupling strength $K=2.2$ corresponding to the on-off intermittency (the blue circles in Fig.~\ref{fig5}~(a),(c)). Curve-fitting using the least squares method allows to prove that the distribution $N(\tau)$ is well-approximated by the dependence $N(\tau)=\alpha \tau^{-3/2}$ (see the red solid lines in Fig.~\ref{fig5}~(a),(c)), which is typical for the on-off intermittency \cite{heagy2}. 

The second intrinsic peculiarity of the observed oscillatory regimes when increasing the coupling strength in the area of the  on-off intermittency consists in the functional dependence of the mean laminar phase length, $<\tau>$, found to be inversely proportional to the critical onset parameter. To visualise this fact, the critical onset parameter introduced in the form $K_2^{\text{crit}}-K$ is used as an argument of the function $<\tau>$. As can be seen from the excellent agreement in Fig.~\ref{fig5}~(b),(d), the dependence of the mean laminar phase length on the critical onset parameter registered in numerical simulations and full-scale experiments is well-approximated by the function $<\tau>= \alpha (K_2^{\text{crit}}-K)^{-1}$ (similarly to Fig.~\ref{fig5}~(a),(c), the least squares method was used). Thus, two functional dependencies being typical for the on-off intermittency \cite{heagy2} were found, $N(\tau) \sim \tau^{-3/2}$ and $< \tau > \sim (K_{2}^{\text{crit}}-K)^{-1}$, which clearly indicates that the observed oscillatory dynamics represents a manifestation of the on-off intermittency. 

\subsection{\label{sec:3.2} Deterministic coherence and anti-coherence resonances}
\begin{figure}[t]
\centering
\includegraphics[width=0.6\textwidth]{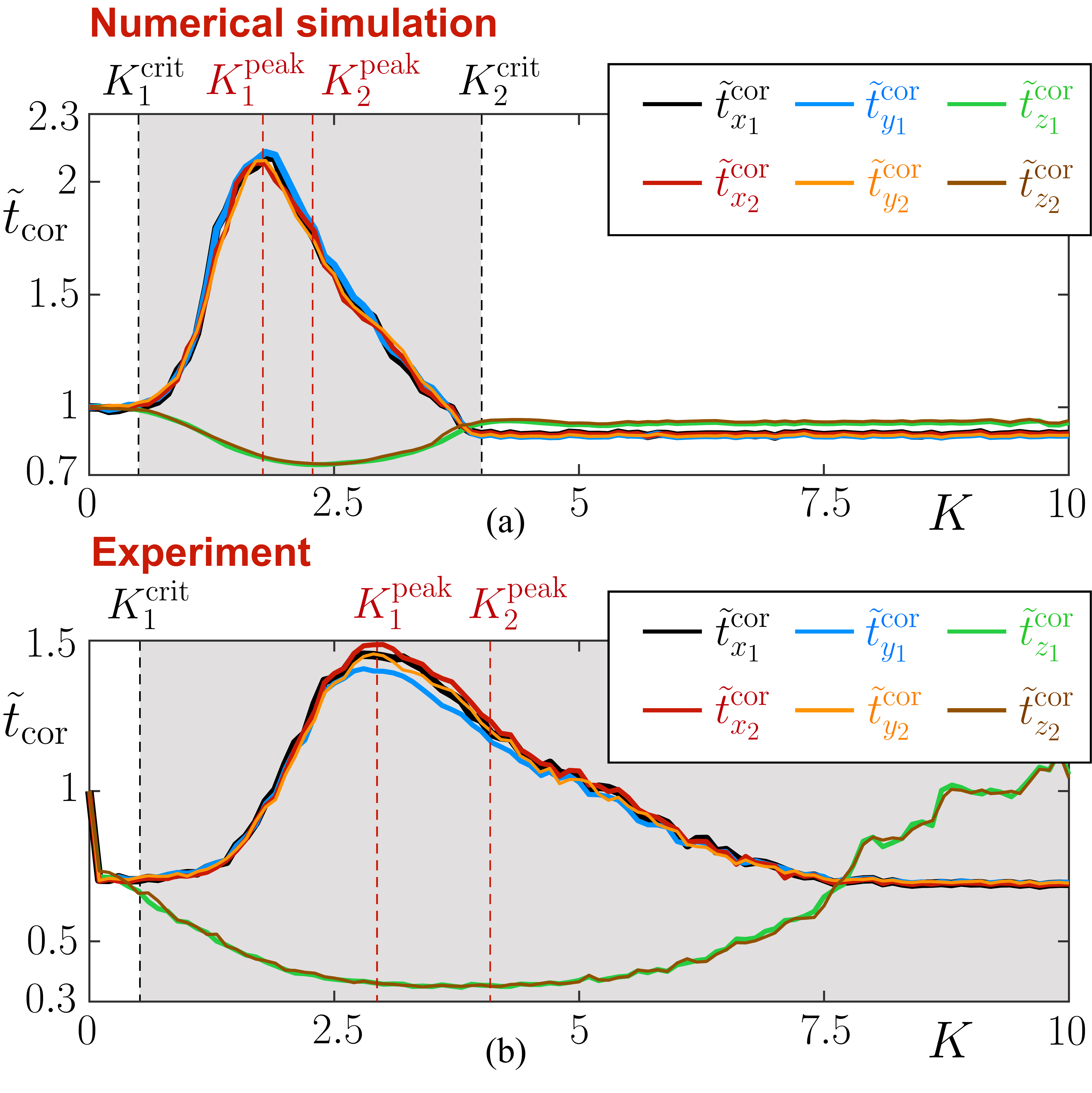}
\caption{Simultaneous occurrence of DCR and DACR when varying the coupling strength $K$ in numerical model (\ref{eq:1}) (panel (a)) and electronic setup (\ref{eq:2}) (panel (b)) highlighted by means of the normalized correlation time $\tilde{t}_{\text{cor}}$. The DCR is manifested as the dependencies $\tilde{t}_{\text{cor}}(K)$, where the normalized correlation time is calculated when analyzing time realizations $x_{1,2}(t)$ and $y_{1,2}(t)$. At the same time, the occurrence of the DACR is reflected in the dependencies $\tilde{t}_{\text{cor}}(K)$ characterizing the evolution of the oscillations $z_{1,2}$. The parameter values of Eqs. (\ref{eq:1}) and (\ref{eq:2}) are $\sigma=10$, $\beta=8/3$, $\rho=28$ (numerical simulation), $P=2.3$ (physical experiment). Random initial conditions within the range [-1:1] are used for numerical simulations. The range $K\in(K^{\text{crit}}_{1}:K^{\text{crit}}_{2})$ corresponds to the occurrence of the on-off intermittency and is coloured in grey.}
\label{fig6}
\end{figure}  
In addition to the exhibition of the on-off intermittency, numerical model (\ref{eq:1}) and electronic setup (\ref{eq:2}) demonstrate DCR and DACR when varying the coupling strength. Intriguingly, both effects occur simultaneously [Fig.~\ref{fig6}]. In particular, increasing $K$ gives rise to the non-monotonic behaviour of the correlation time of  oscillations $x_{1,2}(t)$ and $y_{1,2}(t)$ such that there exists an optimal coupling strength value $K^{\text{peak}}_1\approx 1.8$ (numerical model) and $K^{\text{peak}}_1\approx 2.8$ (experimental setup) corresponding to the most coherent oscillations. Thus, one deals with DCR where the coupling strength plays a role of the noise intensity in classical CR. Meanwhile, analysing oscillations $z_{1,2}$, one can establish the effect of DACR where the non-monotonic dependence of the chaotic oscillation correlation time on the coupling strength results in the local minimum of the oscillation regularity at $K^{\text{peak}}_2\approx 2.3$ (numerical model) and $K^{\text{peak}}_2\approx 3.8$ (experimental setup).

\begin{figure}[t]
\centering
\includegraphics[width=0.9\textwidth]{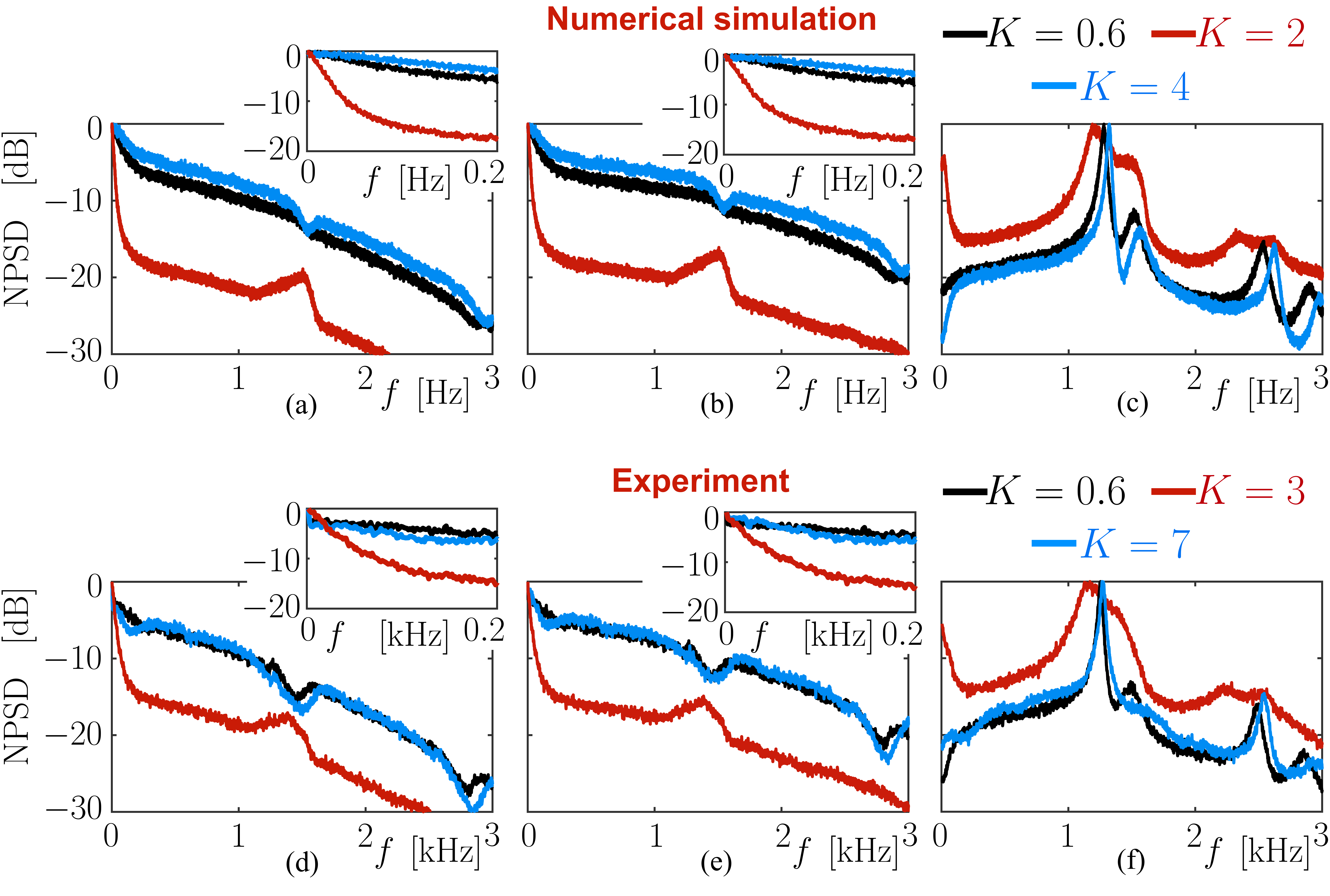}
\caption{Normalized power spectral density (NPSD) evolution when varying the coupling strength $K$ in numerical model (\ref{eq:1}) (panels (a)-(c)) and electronic setup (\ref{eq:2}) (panels (d)-(f)). Panels (a) and (d) depict the NPSD characterising oscillations $x_1(t)$, panels (b) and (e) illustrate the NPSD for oscillations $y_1(t)$, panels (c) and (f) represent the NPSD for oscillations $z_1(t)$. The insets in panels (a), (b), (d), (e) show the NPSD of oscillations $x_1(t)$ and  $y_1(t)$ in the low frequency range. The parameter values of Eqs. (\ref{eq:1}) and (\ref{eq:2}) are $\sigma=10$, $\beta=8/3$, $\rho=28$ (numerical simulation), $P=2.3$ (physical experiment). Random initial conditions within the range [-1:1] are used for numerical simulations. The NPSDs were obtained for three values of the coupling strength: $K=0.6$, $K=2$, $K=4$ (numerical simulation) and $K=0.6$, $K=3$, $K=7$ (physical experiment).}
\label{fig7}
\end{figure}

The nonmonotonic behaviour of the correlation time depicted in Fig. \ref{fig6} fully correlates with the power spectrum evolution when increasing the coupling strength. Analysis of the NPSD allows us to assert that oscillations $x_{1,2}(t)$ and $y_{1,2}(t)$ qualitatively differ from oscillatory processes $z_{1,2}(t)$. In particular, the power spectrum of oscillations $x_{1,2}(t)$ and $y_{1,2}(t)$ are Lorentzian-shaped such that there exists the spectral peak at zero frequency [Fig.~\ref{fig7}~(a),(b),(d),(e)]. The Lorentzian's width first decreases and then increases when the coupling strength growths, which is reflected in the nonmonotonic dependence of the correlation time: a decrease in the full width at half maximum of the power spectrum corresponds to an increase in the correlation time \cite{kubo1991,gardiner1997}. In contrast, the power spectra of processes $z_{1,2}(t)$ are characterized by the spectral peak at nonzero natural frequency [Fig.~\ref{fig7}~(c),(f)]. When the coupling strength increases, the spectral peak becomes wider and then narrows, which fully correlates with the evolution of the correlation time corresponding to coherence resonance \cite{pikovsky,lindner}. 

Both DCR and DACR are exhibited in the range of the coupling strength corresponding to the on-off intermittency area (see the grey area in Fig.~\ref{fig6}). Moreover, the exhibition of both DCR and DACR in numerical model (\ref{eq:1}) ends when the transition to complete synchronization occurs. Such transition could not be realized in electronic setup (\ref{eq:2}) due to experimental restrictions: as noted above, complete synchronization cannot be realized in physical experiments since the available coupling strength range $K\in [0:10]$ is not enough for experimental realization of such regimes. Nevertheless, both DCR and DACR are successfully uncovered in physical experiments and their exhibition is in a good correspondence with results of numerical simulations. It must be noted that the most coherent oscillations $x_{1,2}(t)$ and $y_{1,2}(t)$ are achieved in numerical model (\ref{eq:1}) at $K=K_1^{\text{peak}}$ which corresponds to the local minimum of Lyapunov exponent $\lambda_1$ [Fig.~\ref{fig3}]. Thus, the most regular dynamics corresponds to the lowest values of the maximal Lyapunov exponent, which seems to be a logical and intuitively clear result. 

\section{\label{sec:4} Conclusion}
In the present paper, we report the occurrence of two effects, DCR and DACR, in a system of two interacting Lorenz oscillators. Despite DCR and DACR are in fact the contrary phenomena associated with the existence of local minimum and maximum of the chaotic oscillation regularity, they occur simultaneously when increasing the coupling strength. Analysis of the power spectra and the correlation time evolution has proved the simultaneous exhibition of two effects. Both phenomena were revealed by means of numerical simulation and physical experiments. For the experimental observation of the DCR and DACR, an electronic model of two coupled Lorenz oscillators was developed. The electronic setup demonstrates a good qualitative correspondence to the mathematical model dynamics except of complete synchronization which was not achieved in the coupling strength range available in experiments.

In addition, the considered system exhibits the effect of on-off intermittency interconnected with manifestations of DCR and DACR, since both DCR and DACR are observed in the area of the on-off intermittency exhibition. A similar relationship between on-off intermittency and CR was observed in \cite{liu}, where the existence of the CR phenomenon in on-off intermittency mode was proven both theoretically and numerically for two bidirectionally coupled identical and slightly non-identical Lorenz oscillators. Our numerical studies show that the tendency discovered in \cite{liu} is also valid for the DCR and DACR phenomena that are observed in exactly the same dynamical systems, but without noise. It is important to note that the oscillatory dynamics in the regime of on-off intermittency observed in numerical and physical experiments qualitatively and quantitatively replicates the classical on-off intermittency which is reflected in statistics of the laminar phases \cite{heagy2}. At this moment, the theoretical reasons for the occurrence of the DCR and DACR are not clear and represent an interesting subject for further studies. A comparative analysis of the current results with materials of papers \cite{pisarchik1,jaimes} addressing the issue of DCR and DACR on an example of the coupled Rössler oscillators, allows to conclude that the property of hyperbolicity has no impact on the DCR and DACR (the Rössler attractor is nonhyperbolic, whereas the Lorenz attractor is hyperbolic for the standard parameters, for instance, see the references in paper \cite{semenova2015}) as well as the complex coupling topology (two bidirectionally coupled oscillators are much easier to implement as compared to the star-ring network considered in Refs. \cite{pisarchik1,jaimes}). Moreover, the parameter mismatch taken into consideration in Refs. \cite{pisarchik1,jaimes} is not a principal factor for the observation of DCR and DACR. 

\section*{Declaration of Competing Interest}
The authors declare that they have no known competing financial interests or personal relationships that could have appeared to influence the work reported in this paper.

\section*{Data Availability}
The data that support the findings of this study are available from the corresponding author upon reasonable request.

\section*{Acknowledgements}
P.S.K. and S.V.G. acknowledge support by the Russian Science Foundation (Project No 23-79-30027).

%

\end{document}